\documentclass[pra,aps,twocolumn,superscriptaddress,showpacs]{revtex4}
\usepackage{epsfig,amsmath}

\begin{document}

\title
{Multi-Component Bell Inequality and its
Violation for Continuous Variable Systems}

\author{Jing-Ling Chen}
\email{phycj@nus.edu.sg}
\affiliation{Department of Physics, National University of
Singapore, 2 Science Drive 3, Singapore 117542}
\author{Chunfeng Wu}
\affiliation{Department of Physics, National University of
Singapore, 2 Science Drive 3, Singapore 117542}
\author{L. C. Kwek}
\affiliation{Department of Physics, National University of
Singapore, 2 Science Drive 3, Singapore 117542}
\affiliation{Nanyang Technological University, National Institute
of Education, 1, Nanyang Walk, Singapore 637616}
\author{D. Kaszlikowski}
\affiliation{Department of Physics, National University of
Singapore, 2 Science Drive 3, Singapore 117542}
 \author{M.
\.Zukowski } \affiliation{Instytut Fizyki Teoretycznej i
Astrofizyki, Uniwersytet Gda\'nski, PL-80-952, Gda\'nsk, Poland}
\author{C. H. Oh}
\affiliation{Department of Physics, National University of
Singapore, 2 Science Drive 3, Singapore 117542}

\begin{abstract}
Multi-component correlation functions are developed by utilizing d-outcome measurements. Based on the multi-component correlation functions, we propose a Bell inequality for bipartite d-dimensional systems. Violation of the Bell
inequality for continuous variable (CV) systems is investigated. The
violation of the original Einstein-Podolsky-Rosen state can
exceed the Cirel'son bound, the maximal violation is 2.96981. For finite value of squeezing parameter, violation strength of CV states increases with dimension d. Numerical results show that the violation strength of CV states with finite squeezing parameter is stronger than that of original EPR state.
\end{abstract}

\maketitle

\section{introduction}
In their famous paper of 1935
\cite{EINSTEIN}, Einstein, Podolsky and Rosen (EPR) questioned the
completeness of quantum mechanics, based on a gedanken experiment
involving the position and momentum of two entangled particles.
Einstein believed that there must be elements of reality that
quantum mechanics ignores. It is argued that the incomplete
description could be avoided by postulating the presence of hidden
variables that permit deterministic predictions for microscopic
events. Furthermore, hidden variables could eliminate concerns for
nonlocality. For a long time, EPR argument remained a
philosophical debacle on the foundation of quantum mechanics. In
1964, John Bell made an important step forward in this direction
by considering a version based on the entanglement of spin-$1/2$
particles introduced by Bohm. Bell \cite{Bell} showed that the
assumption of local realism had experimental consequences, and was
not simply an appealing world view. In particular, local realism
implies constraints on the statistics of two or more physically
separated systems. These constraints, called Bell inequalities,
can be violated by the statistical predictions of quantum
mechanics.

Although most of the concepts in quantum information theory were initially developed for quantum systems with discrete variables, many quantum information protocols of continuous variables have also been proposed \cite{protocol}. In recent years, quantum nonlocality for position-momentum variables associated with original EPR states has attracted much attention. The original EPR state is a common eigenstate of the relative
position $\hat{x}_1-\hat{x}_2$ and the total linear momentum
$\hat{p}_1+\hat{p}_2$ and can be expressed as a $\delta$-function:
\begin{eqnarray}
\Psi(x_1,x_2)=\int_{-\infty}^\infty e^{(2\pi
i/\hbar)(x_1-x_2+x_0)p} dp. \label{eq1}
\end{eqnarray} It is important to choose the
appropriate type of observables for testing nonlocality for a
given state. Bell has presented a local realistic model for
position and momentum measurements based on spin-1/2 particles.
He has also argued that the original EPR state would not exhibit
nonlocal effects since the Wigner function representation of the
original EPR state is positive everywhere and therefore admits a
local hidden variable model. Recently, Banaszek and
W\'{o}dkiewicz \cite{BW}  invoked the notion of parity as the
measurement operator and interpreted the Wigner function as a
correlation function for these parity measurements. They then
showed that the EPR state and the two-mode squeezed vacuum state
do not have a local realistic description in the sense that they
violate Bell inequalities such as the Clauser and Horne
inequality \cite{CH} and the Clauser-Horne-Shimony-Holt
(CHSH)\cite{CHSH} inequality. The starting point of the demonstration in \cite{BW} is that the
two-mode Wigner function can be interpreted as a correlation function for the
joint measurement of the parity operator. In the limit $r\rightarrow \infty$, when
the original EPR state is recovered, a significant violation of
Bell inequality takes place, however, the violation is not very
strong. To avoid the unsatisfactory feature, Chen {\it et al.} \cite{ZBCH}
introduced ``pseudospin" operators based on parity, due to the fact that the degree of quantum nonlocality that we can uncover crucially depends not only on the given quantum state but also on the ``Bell operator" \cite{Bell-op}. From reference \cite{ZBCH}, the violation of CHSH inequality for the original EPR state can
reach the Cirel'son bound $2\sqrt{2}$. 

In this paper, we propose a Bell inequality, which is based on multi-component correlation functions, for bipartite systems by utilizing d-outcome measurements. We then investigate violation of the inequality for continuous variable systems. Due to the considered d-outcome measurements, violation of original EPR state can exceed Cirel'son's bound, the maximal vioaltion is 2.96981. The CV case with finite value of squeezing parameter is also studied. The violation strength of CV states with finite squeezing parameter is stronger than that of original EPR state. 

\section{Bell Inequality for
Multi-Component Correlation Functions}

We consider a Bell-type scenario: two space-separated observers, denoted by Alice and Bob, measure two different local observables of d outcomes, labelled by $0, 1, ..., N (=d-1)$. We denote $X_i$ the observable measured by party $X$ and $x_i$ the outcome with $X=A, B (x=a, b)$. If the observers decide to measure $A_1, B_2$, the result is $(0, 4)$ with probability $P(a_1=0, b_2=4)$. Then let us introduce $d$ $N$-dimensional unit vectors 
\begin{widetext}
\begin{eqnarray}
&& {\bf v}_0 = (1,0,0,0,\cdots,0, 0)\nonumber\\
&&{\bf v}_1 = \biggr(-\frac{1}{N},\frac{\sqrt{N^2-1}}{N},0,0,\cdots,0, 0 \biggr)\nonumber\\
&&{\bf v}_2 =  \biggr(-\frac{1}{N},-\frac{1}{N} \sqrt{\frac{(N+1)N}{N(N-1)}},\frac{N-2}{N} \sqrt{\frac{(N+1)N}{(N-1)(N-2)}},0,\cdots,0, 0 \biggr)\nonumber\\
&&\vdots \nonumber\\
&&{\bf v}_{N-1} =  \biggr(-\frac{1}{N},-\frac{1}{N}
\sqrt{\frac{(N+1)N}{N(N-1)}},-\frac{1}{N}
\sqrt{\frac{(N+1)N}{(N-1)(N-2)}},\cdots, -\frac{1}{N}
\sqrt{\frac{(N+1)N}{3\cdot 2}},\frac{1}{N} \sqrt{\frac{(N+1)N}{2\cdot 1}} \biggr)\nonumber\\
&&{\bf v}_{N} =  \biggr(-\frac{1}{N},-\frac{1}{N}
\sqrt{\frac{(N+1)N}{N(N-1)}},-\frac{1}{N}
\sqrt{\frac{(N+1)N}{(N-1)(N-2)}},\cdots, -\frac{1}{N}
\sqrt{\frac{(N+1)N}{3\cdot 2}},-\frac{1}{N}
\sqrt{\frac{(N+1)N}{2\cdot 1}} \biggr)
\end{eqnarray}
\end{widetext}
These $d$ vectors satisfy following properties:
\begin{eqnarray}
&& (i) \;\;\;\;\;  \sum_{j=0}^{N} {\bf v}_j =0 \nonumber\\
&& (ii) \;\;\;\;\; {\bf v}_j \cdot {\bf v}_k \equiv -\frac{1}{N}
\;\;\;\; (j\ne k)
\end{eqnarray}
For $d=2$, it is just two valued variables (i.e., $v_0=1, v_1=-1$)
obtained from a measurement. If the measure result of Alice is $m$, and Bob's result is
$n$ (where $m$ and $n$ are less than $N$), we then associate a vector ${\bf v}_{m+n}$ for the
correlation between Alice and Bob [${\bf v}_{m+n}$ understood as
${\bf v}_t$, where $t=(m+n), {\rm mod}\; d$]. Based on which, we
now construct multi-component correlation functions:
\begin{eqnarray}
{\vec Q}_{ij} & = & \sum_{m, n} {\bf v}_{m+n} P(a_i=m,b_j=n)
\nonumber \\ & = & \sum_{t=0}^{d}{\bf v}_t P(m+n=t)
\end{eqnarray}
where $P(a_i=m,b_j=n)$ is the joint probability of $a_i$ obtain
outcome $m$ and $b_j$ obtain outcome $n$ , and ${\vec Q}_{ij}=
(Q_{ij}^{(0)},Q_{ij}^{(1)},Q_{ij}^{(2)},\cdots,Q_{ij}^{(N-1)})$,
$Q_{ij}^{(k)}$ represents the $k$-th component of the vector
correlation function ${\vec Q}_{ij}$.

We now define a Bell quantity for the multi-component
correlation functions,  
\begin{eqnarray}
 {\cal B}_d & = & {\cal B}^{(0)} + \sqrt{\frac{N(N-1)}{(N+1)N}} {\cal
B}^{(1)} + \sqrt{\frac{(N-1)(N-2)}{(N+1)N}} {\cal B}^{(2)} \nonumber \\
&&+\cdots + \sqrt{\frac{2\cdot 1}{(N+1)N}} {\cal B}^{(N-1)} \nonumber\\
 &=& \sum_{k=0}^{N-1} \sqrt{\frac{(N+1-k)(N-k)}{(N+1)N}} {\cal
 B}^{(k)}
\end{eqnarray}
where
\begin{eqnarray}
{\cal B}^{(0)} &=&
Q_{11}^{(0)}+Q_{12}^{(0)}-Q_{21}^{(0)}+Q_{22}^{(0)}\nonumber\\
{\cal B}^{(k)}&=&
Q_{11}^{(k)}-Q_{12}^{(k)}-Q_{21}^{(k)}+Q_{22}^{(k)} ,\;\;\;\;
(k\ne 0)
\end{eqnarray}
Any local realistic description of the previous Gedanken experiment imposes the following inequality:
\begin{eqnarray}
&& -2\biggr(\delta_{2d}+(1-\delta_{2d}) \frac{d+1}{d-1}\biggr) \le
{\cal B}_d  \le 2
\end{eqnarray}
Obviously, this inequality reduces to the usual CHSH inequality
for $d=2$.

The quantum prediction for the joint probability reads
\begin{eqnarray}
P^{QM}(a_i=m,b_j=n)= \langle \psi|
\hat{P}(a_i=m)\otimes\hat{P}(b_j=n)|\psi\rangle
 \label{eq2}
\end{eqnarray}
where $i, j=1,2$; $m,n=0,...,N$, $\hat{P}(a_i=m)={\cal U}_A^{\dagger}|m\rangle\langle m|{\cal
U}_A$ is the projector of Alice for the $i$-th measurement and similar
definition for $\hat{P}(b_j=n)$. 

It is well known that the two-mode squeezed vacuum state can be
generated in the nondegenerate optical parametric amplifier
(NOPA)\cite{NOPA} is

\begin{equation}
| {\rm NOPA}\rangle =e^{r(a_1^{\dagger }a_2^{\dagger
}-a_1a_2)}|00\rangle =\sum_{n=0}^\infty \frac{(\tanh r)^n}{\cosh r}%
|nn\rangle ,  \label{nopa}
\end{equation}
where $r>0$ is the squeezing parameter and $\left| nn\right\rangle
\equiv \left| n\right\rangle _1\otimes \left| n\right\rangle _2=\frac 1{n!}%
(a_1^{\dagger })^n(a_2^{\dagger })^n\left| 00\right\rangle $. The
NOPA states $\left| {\rm NOPA}\right\rangle $ can also be written
as \cite{BW}:
\begin{eqnarray}
\left| {\rm NOPA}\right\rangle =\sqrt{1-\tanh ^2r}\int dq\int
dq^{\prime }g(q,q^{\prime };\tanh r)|qq^{\prime }\rangle , 
\label{epr} \nonumber
\end{eqnarray}
where $g\left( q,q^{\prime };x\right) \equiv \frac 1{\sqrt{\pi
(1-x^2)}}\exp
\left[ -\frac{q^2+q^{\prime 2}-2qq^{\prime }x}{2(1-x^2)}\right] $ and $%
\left| qq^{\prime }\right\rangle \equiv \left| q\right\rangle
_1\otimes \left| q^{\prime }\right\rangle _2$, with $\left|
q\right\rangle $ being the usual eigenstate of the position
operator. Since $\lim_{x\rightarrow
1}g\left( q,q^{\prime };x\right) =\delta (q-q^{\prime })$, one has $%
\lim_{r\rightarrow \infty }\int dq\int dq^{\prime }g(q,q^{\prime
};\tanh r)|qq^{\prime }\rangle =\int dq|qq\rangle =\left| {\rm
EPR}\right\rangle $, which is just the original EPR state. Thus,
in the infinite squeezing limit, $\left| {\rm NOPA}\right\rangle
\left| _{r\rightarrow \infty }\right. $ becomes the original,
normalized EPR state. Following Brukner {\it et al.} \cite{Brukner}, we can map the two-mode squeezed state onto a $d$-dimensional pure state:
\begin{eqnarray}
&& |\psi_d\rangle =\frac{{\rm sech}
r}{\sqrt{1-\tanh^{2d}r}}\sum_{n=0}^{d-1}(\tanh r)^n |nn\rangle.
\end{eqnarray}
If the measurement result of Alice is $m$ photons, and Bob's result is
$n$ photons, we then ascribe a vector ${\bf v}_{m+n}$ for the
correlation between Alice and Bob. And $P(a_i=m,b_j=n)$ is the joint probability of $a_i$ obtain
$m$ photons and $b_j$ obtain $n$ photons. More precisely, for the two-mode squeezed state one obtains following joint probability
\begin{eqnarray}
P^{QM}(a_i=m,b_j=n)= \langle \psi_d|
\hat{P}(a_i=m)\otimes\hat{P}(b_j=n)|\psi_d\rangle
\end{eqnarray}

\section{Some Examples}

For $d=3$, we have three outcomes ${\bf v}_0=(1,0)$, ${\bf
v}_1=(-1/2, \sqrt{3}/2)$, ${\bf v}_2=(-1/2,-\sqrt{3}/2)$. Accordingly the NOPA state is divided into
three groups, namely,
\begin{eqnarray}
|{\rm NOPA}\rangle  & = & \frac{1}{\cosh r}\sum^{\infty}_{n=0}
\biggr(\tanh^{3n}r|3n\rangle|3n\rangle \nonumber\\
& & +\tanh^{3n+1}r|3n+1\rangle|3n+1\rangle \nonumber \\
& & +\tanh^{3n+2}r|3n+2\rangle|3n+2\rangle\biggr)
\end{eqnarray}
The two-component correlation functions depend on quantum version of joint probabilities
\begin{eqnarray}
&&P^{QM}(a_i=m,b_j=n) \nonumber \\
 &&=\langle {\rm NOPA}|
\hat{P}(a_i=m)\otimes\hat{P}(b_j=n)|{\rm NOPA}\rangle
\end{eqnarray}

Local realistic description imposes $-4\le {\cal B}\le 2$. Numerical results show that ${\cal B}_{d=3}(r = 1.4068) \simeq
2.90638$; ${\cal B}_{d=3}(r \rightarrow \infty) = 4/(6\sqrt{3}-9)
\simeq 2.87293$. For ${\cal B}_{d=3}(r \rightarrow \infty)$, the
four optimal two-component quantum correlations read:
\begin{eqnarray}
&& \vec{Q}_{11}=\vec{Q}_{22}=\vec{Q}_{12}^*=(\frac{2\sqrt 3 +1}{6},-\frac{2-\sqrt
3}{6}), \nonumber \\ && \vec{Q}_{21}=(-{1\over 3},-{2\over 3}), \nonumber\\
&& |\vec{Q}_{ij}|=\sqrt{(Q^x_{ij})^2+(Q^y_{ij})^2}=\frac{\sqrt 5}{3}.
\end{eqnarray}
One thing worth to note that one can treat those three two-dimensional vectors in
terms of complex numbers, namely, $v_0=1$, $v_1=\omega$ and
$v_2=\omega^2$ for simplicity, where $\omega=\exp[i2\pi/3]$. Now the Bell
inequality becomes \cite{JLC}
\begin{eqnarray}
-4\leq & {\rm Re}[Q_{11}+Q_{12}-Q_{21}+Q_{22}] &\nonumber \\
& +\frac{1}{\sqrt{3}}{\rm Im}[Q_{11}-Q_{12}-Q_{21}+Q_{22}] &\leq2.
\end{eqnarray}
In this sense, we can check the generalized ``parity" operator
$(\omega)^{\hat{n}}$ other than usual parity operator $(-1)^{\hat{n}}$, the two-component correlation function also reads
\begin{eqnarray}
\mbox{\hspace{-0.8cm}}&& Q_{ij}=\langle {\rm NOPA}|{\cal
U}_A^{\dagger}\otimes {\cal U}_B^\dagger \; (\omega)^{{\hat
n}_a+{\hat n }_b} \; {\cal U}_A \otimes {\cal U}_B |{\rm
NOPA}\rangle
\end{eqnarray}
where ${\cal U}_{A,B}$ are generally $U(3)$ transformations and we
can sufficiently take them as the products of three spin-coherent
operators:
\begin{eqnarray}
{\cal U}_A=e^{\xi_3 \hat{U}_+ - \xi_3^*\hat{U}_-} \; e^{\xi_2
\hat{V}_+ - \xi_2^*\hat{V}_-}\; e^{\xi_1 \hat{I}_+ -
\xi_1^*\hat{I}_-} \nonumber
\end{eqnarray}
where $\xi_j=\frac{\theta_j}{2}e^{-i\varphi_j}$ [actually, the
phases $\varphi_j$ can be set to be zero, since they do not affect
the maximal violation. Hence ${\cal U}_{A,B}$ is a $SO(3)$
rotation]. $\hat{I}_\pm$, $\hat{U}_\pm$ and $\hat{V}_\pm$ are
pseudo-$su(3)$-spin which can be realized by the Fock states as
\begin{eqnarray}
&&{\hat I}_+=\sum^{\infty}_{n=0}|3n\rangle\langle3n+1|, \ \ \ \ \
{\hat I}_-=\sum^{\infty}_{n=0}|3n+1\rangle\langle3n|,  \nonumber\\
&& {\hat I}_z=\frac{1}{2}(|3n\rangle\langle3n|-|3n+1\rangle\langle3n+1|), \nonumber \\
&&{\hat U}_+=\sum^{\infty}_{n=0}|3n+1\rangle\langle3n+2|, \ \ \ \
\
{\hat U}_-=\sum^{\infty}_{n=0}|3n+2\rangle\langle3n+1|, \nonumber\\
&& {\hat
U}_z=\frac{1}{2}(|3n+1\rangle\langle3n+1|-|3n+2\rangle\langle3n+2|) \nonumber \\
&&{\hat V}_+=\sum^{\infty}_{n=0}|3n\rangle\langle3n+2|, \ \ \ \ \
{\hat V}_-=\sum^{\infty}_{n=0}|3n+2\rangle\langle3n|,   \nonumber\\
&& {\hat
V}_z=\frac{1}{2}(|3n\rangle\langle3n|-|3n+2\rangle\langle3n+2|). \nonumber
\end{eqnarray}
Operators $\{{\hat I}_\pm,{\hat I}_z \}$, $\{{\hat U}_\pm,{\hat
U}_z \}$ and $\{{\hat V}_\pm,{\hat V}_z \}$ form three $SU(2)$
groups, respectively, and $\{ {\hat I}_\pm, {\hat U}_\pm, {\hat
V}_\pm,  {\hat I}_z, ({\hat U}_z+{\hat V}_z)/\sqrt{3}\}$ forms a
$SU(3)$ group.

\begin{figure}
\begin{center}
\epsfig{figure=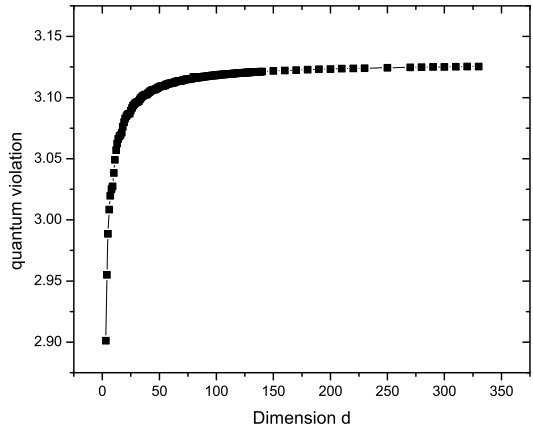,width=0.44\textwidth}
\end{center}
\caption{Violation of multi-component Bell inequality for CV states with finite squeezing parameter for different d.} \label{fig}
\end{figure}

We can similiarly get ${\cal B}_{d}(r = {\rm finite} \ {\rm value})$ and ${\cal B}_{d}(r \rightarrow \infty)$ with different d. We list them partly in Table \ref{tab}. Obviously, the degree of the violation increases with dimension d, and the violation strength of CV states with finite squeezing parameter is stronger than that of original EPR state. When squeezing parameter goes to infinity, i.e., original EPR state gotten, we can find the bound of the violation of multi-component correlation-function Bell inequality \cite{CGLMP}. 
\begin{widetext}
\begin{eqnarray}
\lim_{d \rightarrow \infty} {\cal
B}_{d}(r \rightarrow \infty) &=& \lim_{d \rightarrow \infty}4d \sum_{k=0}^{[d/2]-1}(1-\frac{2k}{d-1})(\frac{1}{2d^3\sin ^2[\pi(k+\frac{1}{4})/d]}-\frac{1}{2d^3\sin ^2[\pi(-k-1+\frac{1}{4})/d]}) \nonumber \\
& =& \frac{2}{\pi^2}
\sum_{k=0}^{\infty}[\frac{1}{(k+1/4)^2}-\frac{1}{(k+3/4)^2}]
\simeq 2.96981
\end{eqnarray}
\end{widetext}
It is interesting to note that for ${\cal
B}_{d}(r\rightarrow \infty)$, the four optimal multi-component
quantum correlations share the same module: $|{\vec
Q}_{ij}|=\sqrt{\frac{2d-1}{3d}}$. When $d$ tends to infinity,
$|{\vec Q}_{ij}|=\sqrt{2/3}$. However, we do not have an analytical way to find a bound for violation with finite squeezing parameter. For this case, what we do is draw a graph to see the variation of ${\cal B}_{d}(r = {\rm finite} \ {\rm value})$ with increasing dimension, see Fig.\ref{fig}. We calculate the maximal quantum violation for CV states with different d. The more the value of dimension, the more difficult to find a maximal violation. Hence the violation strength points we get are for $d\leq 330$. With these values, it is easy to see that the violation increases from slowly to slowly with increasing of d. Which means that there exists a limit for quantum violation when d goes to infinity. Until now, we do not have an exact value of the limit. We use a software, which can give experimence expression given enough points, to find a expression that describes the curve in Fig.\ref{fig},
\begin{eqnarray}
B=3.12885-1.06535/d+2.13122/d^2-2.19262 e^{-d}
\end{eqnarray}
When $d\rightarrow \infty$, quantum violation (B) goes to 3.12885. Hence, such value can be thought as an approximate violation limit for CV states with finite squeezing parameter.

\begin{table}
\begin{tabular}{|c|c|c|c|c|c|c|}
     \hline\hline
$\langle {\cal B}_{d}\rangle $
     &   $d=5$       &  $d=10$       &  $d=15$       &  $d=20$  &  $d=25$      \\
     \hline
${\rm EPR}$
     &   $2.91055$  &  $2.9398$       &  $2.94973$  &  $2.95473$ &  $2.9577$  \\
     \hline
${\rm NOPA}$
     &   $2.9886$  &  $3.03842$  &  $3.06836$  &  $3.08273$  &  $3.08932$ \\
${\rm with \ r}$
     &   $(1.44614)$  &  $(1.72082)$  &  $(1.8366)$  &  $(1.96562)$  &  $(2.07377)$  \\
\hline\hline
\end{tabular}
\caption{Violation of multi-component Bell inequality for $|{\rm NOPA}\rangle$ and $|{\rm EPR}\rangle$ states with different d.}
\label{tab}\end{table}

\section{Conclusion}
In summary, we construct multi-component correlation functions based on d-outcome measurements. A Bell inequality for bipartite d-dimensional systems is developed accordingly. We then investigate violation of such Bell inequality for continuous variables case. The degree of the violation increases with dimension d, and the limit of the violation for the original EPR state is found to be 2.96981, which exceeds the Cirel'son bound. The reason for this is due to the fact that we consider d($>2$)-outcome measurements. Numerical results show that the violation strength of CV states with finite squeezing parameter is stronger than that of original EPR state.

This work is supported by NUS academic research grant WBS: R-144-000-089-112. J.L.C acknowledges financial support from Singapore Millennium Foundation and (in part) by NSF of China (No. 10201015).

{\it Note added:} While completing this work we learn a similar
result obtained in Ref. \cite{SLK}, which based on the CGLMP
inequality \cite{CGLMP}.

\end{document}